\documentstyle[prd,preprint,tighten,aps,eqsecnum,%
amssymb,newlfont]{revtex}
\newcommand{\be}{\begin{equation}}
\newcommand{\ee}{\end{equation}}
\newcommand{\ba}{\begin{eqnarray}}
\newcommand{\ea}{\end{eqnarray}}
\newcommand{\no}{\nonumber}
\newcommand{\n}{\!\not\!}
\begin{document}
\draft
\preprint{
\begin{tabular}{r}
UWThPh-1997-20\\
August 1997
\end{tabular}
}
\title{Effects of neutrino oscillations and neutrino magnetic
moments on elastic neutrino--electron scattering}
\author{W. Grimus and P. Stockinger}
\address{Institute for Theoretical Physics, University of Vienna,\\
Boltzmanngasse 5, A--1090 Vienna, Austria.}
\maketitle
\begin{abstract}
We consider elastic $\bar\nu_e e^-$ scattering taking into account
possible effects of neutrino masses and mixing and of neutrino
magnetic moments and electric dipole moments. Having in mind
antineutrinos produced in a nuclear reactor we compute, in particular, 
the weak--electromagnetic interference terms which are linear
in the magnetic (electric dipole) moments and also in the neutrino
masses. We show that these terms are, however, suppressed compared to
the pure weak and electromagnetic cross sections. We also comment upon
the possibility of using the electromagnetic cross section to
investigate neutrino oscillations.
\end{abstract}


\narrowtext

\section{INTRODUCTION}

Exact knowledge of the intrinsic neutrino properties is 
decisive for our understanding of fundamental interactions and 
if and how the standard model has to be extended in the lepton sector. 
So far, intensive experimental efforts have been made in neutrino
oscillations \cite{pon} to pin down neutrino masses and mixing angles
\cite{boehm}. As for neutrino magnetic moments ($\mu$) and 
electric dipole moments
($d$), elastic neutrino--electron scattering is a simple and fundamental
process with great sensitivity for these quantities. This process has been
used to obtain the experimental limits
$\mu_{\nu_e}<1.8 \times 10^{-10}\mu_B$ \cite{derbin} and
$\mu_{\nu_{\mu}}<7.4 \times 10^{-10}\mu_B$ \cite{krakauer} where $\mu_B$ is
the Bohr magneton. A much greater sensitivity for $\mu_{\nu_e}$ is planned
in the MUNU experiment \cite{munu1,munu2} 
which makes use of reactor neutrinos like
the experiments evaluated in Ref. \cite{derbin}. 
The relevant cross section for elastic
$\bar\nu_e e^-$ scattering with $\mu_{\nu_e} \neq 0$ is given by \cite{engl}
\be
\label{wq}
\frac {d \sigma}{d T} = \frac{d \sigma_w}{d T}+ \frac {d\sigma_m}
{dT}
\ee
where\footnote{For $\nu_e e^-$ scattering $G_A$ has to be replaced by $-G_A$.}
\be\label{wq1}
\frac{d \sigma_w}{d T}= \frac {G_F^2 m_e} {2\pi} \left[ (G_V -G_A)^2
+ (G_V+G_A)^2 (1- \frac{T}{E_\nu})^2 - (G_V^2- G_A^2) \frac{m_eT}{E_\nu^2}
\right] 
\ee
and
\be\label{wq2}
\frac{d \sigma_m}{d T}= \frac {\alpha^2 \pi} {m_e^2} (\mu'_{\nu_e})^2
\left( \frac{1}{T} -\frac{1}{E_\nu}\right) 
\ee
with $(\mu'_{\nu_e})^2 = (\mu'^2 + d'^2)_{\nu_e}$.
In these formulas, neutrino masses and mixing have not been taken into account
and the electron neutrino is assumed to be a Dirac particle.
Furthermore,
$T=E_e ' - m_e$ is the kinetic energy of the recoil electron,
magnetic and electric dipole moments are given in units of the Bohr 
magneton, i.e. $\mu' \equiv \mu/\mu_B$ and $d' \equiv d/\mu_B$, 
$G_F$ is the Fermi coupling constant, $\alpha \approx 1/137 $ is the fine 
structure constant and $G_{V,A} \equiv g_{V,A} + 1$ with $g_V =
2 \sin^2 \theta_W -1/2$ and $g_A = - 1/2$.
$g_{V,A}$
correspond to the neutral current contribution to the effective 
interaction Lagrangian of this process.\par 
In Eq. (\ref{wq}) 
there is no interference between the weak and electromagnetic interaction.
This is evident since for negligible neutrino mass
the electromagnetic interaction involving magnetic and
electric dipole moments leads
to a helicity flip of the neutrino while the weak interaction always 
conserves helicity.
Such an interference term is, however, only absent
as long as the incident neutrino flux has no transverse polarization 
\cite{barb}.
Also possible scalar exchange leads to an interference with the
neutrino magnetic or electric dipole moment interaction
\cite{stock}. Moreover, a weak--electromagnetic interference term will appear 
if we do not
neglect neutrino masses. The subject of our 
investigation will be the effect of neutrino masses and mixing on
elastic neutrino--electron scattering with magnetic and electric
dipole moments of neutrinos.

If neutrinos are massive particles, the mass matrix in the weak basis
of the neutrino fields is in general non-diagonal. 
After diagonalization, the left-handed neutrino fields in the weak basis 
are linear combinations of mass eigenfields, i.e.
\be
\label{lcom}
\nu_{\alpha L}(x)=U_{\alpha j}\nu_{jL}(x)
\ee
where $U$ is a unitary matrix, the neutrino fields $\nu_j$ have masses
$m_j$ and the index $\alpha$ indicates the flavours $e,\mu$ and $\tau$. 
Eq. (\ref{lcom}) gives rise to the phenomenon of neutrino oscillations.
Since we want to incorporate effects linear in the neutrino masses $m_j$
into elastic $\bar\nu_e e^-$ scattering we are looking for
weak--electromagnetic interference terms proportional to $m_\nu \times
\mu_\nu(d_\nu)$
in the cross section ($m_\nu$ is a generic neutrino mass and
$\mu_\nu$ $(d_\nu)$ is a generic neutrino magnetic (electric dipole) moment).
It is therefore necessary to know the initial neutrino state up to
terms linear in the neutrino masses. It has been stressed in the literature
\cite{rich} that in order to have control over the initial 
state a safe way is 
to consider neutrino production and detection as a single process.\par
In the present paper we will proceed along this line. We will show
that under certain conditions the production process, which is 
$\beta$-decay of fission products in a reactor in our case, factorizes from
elastic neutrino--electron scattering in an unambiguous way.
We will do the calculation for both Dirac and Majorana neutrinos and 
finally we will discuss the $m_\nu \times \mu_\nu(d_\nu)$ interference term
and also the effects of neutrino mixing on Eqs. (\ref{wq}-\ref{wq2}).

\section{The Amplitude}

We consider the following combined production--detection process
\ba \label{process}
^A_ZX \rightarrow \; ^{A}_{Z+1} X'+e_S^- +\bar\nu & & \no \\
& \searrow & \no \\
& & \bar\nu +e^-\rightarrow\bar\nu +e^-
\ea
where the neutrinos are produced by $\beta$-decay of fission nuclei $^A_ZX$
in a reactor and 
detected by elastic scattering off electrons. The initial particles $^A_ZX$ 
and $e^-$ (detector electron) are localized and macroscopically separated by a 
distance $L$. With $e^-_S$ we denote the 
electron which is generated in the $\beta$-decay and absorbed in the reactor.
This is exactly the 
situation described in Ref. \cite{grim} and hence it is possible to use the 
methods and results obtained there. In this paper we will concentrate
on the interference term between weak and electromagnetic 
interaction and calculate all terms in the cross section which depend 
linearly on the neutrino masses.

The weak and electromagnetic 
interaction Langrangians relevant for the detection are
\be \label{Lagw}
\mathcal{L}_{w}(x)=-\frac{G_F}{\sqrt{2}} \sum_{j,k} 
      \bar e(x) \gamma^\mu(Z^V_{jk}-Z^A_{jk}\gamma_5)e(x) \,
      \bar \nu_j(x) \gamma_\mu (1-\gamma_5)\nu_k(x)
\ee
with
\be
Z^{V,A}_{jk}\equiv U^*_{ej}U_{ek}+\delta_{jk}g_{V,A}
\ee
and
\be \label{Lagm}
\mathcal{L}_{em}(x)=
     -\frac{1}{2} \sum_{j,k}\bar\nu_j(x)\left(\mu_{jk}+id_{jk}
     \gamma_5\right)\sigma_{\lambda\rho}\nu_k(x)F^{\lambda\rho}(x)
     +e\, \bar e(x)\gamma^\lambda e(x) A_\lambda(x) \, ,
\ee
respectively, where $A_\lambda$ is the electromagnetic vector potential,
$F^{\lambda\rho}$ the field strength tensor, $e$ the charge of the positron,
$P_{R,L} = (1\pm \gamma_5)/2$ and the $\mu_{jk}$ $(d_{jk})$ are the magnetic
(electric dipole) moments and transition moments (for $j\not= k$) of 
the mass eigenfields. For Dirac neutrinos, the amplitude 
for the process (\ref{process}) in the limit of a macroscopic distance L
with an antineutrino
of mass $m_\ell$ in the final state is given by \cite{grim}
\ba
\label{amp}
\mathcal{A}_\ell & = & \sum_j 
\tilde J_\lambda\bar u_{eS}\gamma^\lambda P_L(-\n k_j
+m_j)e^{iq_jL} U_{ej} \no\\
& & \times\bigg\{ \sqrt{2} G_F \gamma^\mu P_L v(k'_\ell) \:
\bar u_e(p')\gamma_\mu
( Z^V_{j\ell} - Z^A_{j\ell} \gamma_5 ) \tilde\psi_e\no\\
& & -ie (\mu_{j\ell}+id_{j\ell}\gamma_5) q_\mu \sigma^{\mu\nu}v(k'_\ell) \:
\frac{g_{\nu \rho}}{q^2} \:
\bar u_e(p')\gamma^\rho\tilde\psi_e\bigg\}
\ea
where $\tilde{J}_\lambda$
is the Fourier transform of the hadronic current,
$\tilde{\psi}_e(\vec{p})$ is the Fourier transform of 
the wave function $\psi_e$
of the initial detector electron, $u_{eS}$ denotes 
the spinor of the source electron,
$p$ and $p'$ the
initial\footnote{$p=(\sqrt{m_e^2+\vec{p}\,{}^2},\vec{p}\, )$} and final
momenta, respectively, of the detector electron, 
$k_j$ $(k'_\ell)$ the momenta of the 
intermediate (final) neutrinos and
\be 
q=p-p', \quad k_j = {E_\nu \choose q_j\vec{l}}, \quad q_j = \sqrt{E_\nu^2
- m_j^2}
\ee
where $\vec{l}$ is the unit vector pointing from the neutrino source to
the detection point.
Apart from the factor $\exp (iq_jL)$, the amplitude (\ref{amp}) in
the approximation of macroscopic separation of the neutrino source and the
detector is just the sum over the products of 
production and detection amplitude of antineutrinos with mass $m_j$
\cite{grim}.\par 

\noindent The definitions
\be
M_{j\ell}= -i (\mu_{j\ell}+ id_{j\ell}\gamma_5) q_\mu L_\nu^{em}\sigma^
{\mu\nu}\; , 
\ee
\be
W_{j\ell}= \gamma^\mu P_L L_{\mu j\ell} 
\ee 
with
\be
L_\nu^{em} = \frac{e}{q^2} \bar{u}_e(p') \gamma_\nu \tilde{\psi}_e
\ee
and
\be
L_{\mu j\ell} = \sqrt{2} G_F \bar u_e(p') \gamma_\mu 
(Z^V_{j\ell}-Z^A_{j\ell}\gamma_5)\tilde\psi_e
\ee
will serve in the following to simplify our notations as far as possible and 
to easily keep track of the interactions involved in the various terms of
the cross section. With the help of these definitions the amplitude
(\ref{amp}) is rewritten as
\be
\label{amp1}
\mathcal{A}_\ell = \sum_j 
\tilde J_\lambda\bar u_{eS}\gamma^\lambda P_L(-\n k_j
+m_j)e^{iq_jL} U_{ej}[W_{j\ell}+M_{j\ell}]v(k'_\ell)\; .
\ee

In the next section we will show that under certain conditions also 
the cross section can be factorized into a product of production and 
detection cross section.

\section {The Cross Section}

In order to simplify the calculations it is necessary to make an 
approximation for the state of the detector electron. 
We will assume henceforth
that the momentum spread of this electron is negligible and thus use
for its description simply a spinor of an electron at rest. 
Summing over the spins of the leptons and over the 
absolute squares of the amplitudes $\mathcal{A}_\ell$ we obtain
\ba
\sum_\ell \sum_{\mathrm{spins}}\mathcal{A}_\ell\mathcal{A}_\ell^*
& = &\mbox{Tr}\left\{\,
\hat{\tilde{\!\!\n J}} \n p_S\n{\tilde{J}} P_L(-\n k_j+m_j)
(W_{j\ell}+M_{j\ell})(\n k'_\ell-m_\ell)(\hat{W}_{\ell n}+\hat{M}_{\ell n}
)(-\n k_n+m_n)\right\}\no\\
& &\times e^{i(q_j-q_n)L}U_{ej}U^*_{en} \label{sa1}
\ea
where
\be
\hat{W}_{\ell n} \equiv \gamma^\mu P_L L^*_{\mu n \ell}, \qquad
\hat{M}_{\ell n}\equiv i(\mu^*_{n\ell}+id^*_{n\ell})q_\mu L^{em*}_\nu 
\sigma^{\mu\nu}
\ee
and
\be
\n \tilde{J}\equiv \tilde J_\lambda\gamma^\lambda,\qquad
\hat{\tilde{\!\!\n J}}\equiv\tilde{J}^*_\lambda\gamma^\lambda\;.
\ee
The momentum of the electron from the source is denoted by $p_S$.

Now we will approximate the right-hand side of Eq. (\ref{sa1})
by confining ourselves to terms which are at most linear 
in the masses of the neutrinos. 
Fixing the spatial neutrino momenta, the neutrino energies 
depend quadratically on the neutrino masses for $m_\nu^2 \ll k_\nu^2$
where $k_\nu$ denotes a generic neutrino momentum. Therefore, linear
dependence on the neutrino masses derives from the explicit mass 
factors $m_{j,\ell,n}$ in Eq. (\ref{sa1}) and the linear approximation
amounts to neglecting all neutrino masses in the neutrino 4-momenta:
\be
k_j=k\;\, \forall j,\qquad k'_\ell=k'\;\,\forall \ell
\ee
with
\be
k^2=k'^2=0.
\ee
Note, however, that we keep the squares of neutrino masses in the
phase factors $e^{i(q_j-q_n)L}$ to allow for neutrino oscillations. 
Eq. (\ref{sa1}) becomes in this approximation
\ba
\label{sa2}
\sum_\ell \sum_{\mathrm{spins}} 
\mathcal{A}_\ell \mathcal{A}_\ell^* & = & e^{i(q_j-q_n)L}U_{ej}
U^*_{en} 
\mbox{Tr} \bigg\{\, \hat{\tilde{\!\!\n J}} \n p_S\n{\tilde{J}} P_L        
\left[ \,\n k(W_{j\ell}+M_{j\ell})\n k'(\hat{W}_{\ell n}+
\hat{M}_{\ell n})\n k \right. \no\\[-3mm]
& & -\n k(W_{j\ell}+M_{j\ell})m_\ell (\hat{W}_{\ell n}+\hat{M}_{\ell n})\n k
-m_j(W_{j\ell}+M_{j\ell})\n k'(\hat{W}_{\ell n}+\hat{M}_{\ell n})\n k\no\\
& & \left. -\n k(W_{j\ell}+M_{j\ell})\n k'(\hat{W}_{\ell n}+\hat{M}_{\ell
n})m_n \right] \bigg\}\, .
\ea

Now we would like to separate neutrino production from neutrino
scattering by representing Eq. (\ref{sa2}) as a product of two factors
describing production and scattering, respectively, conforming with
the usual situation. Actually, this can easily be done for the two
terms in the square brackets in Eq. (\ref{sa2}) which have $\n k$ on
both sides.  Since now neutrinos are 
``massless'' we can write
\be
\label{eq2}
v(k,s_+)\bar v(k,s_+)=P_L\n k
\ee
where $v(k,s_+)$ represents an antineutrino with positive helicity.
Using Eq. (\ref{eq2}) we can split the first two terms of Eq. (\ref{sa2}) into 
a product of production and detection trace and obtain for these terms
\ba
\label{eq3}
& & e^{i(q_j-q_n)L} U_{ej}U^*_{en} \,
\mbox{Tr} \left\{\,\hat{\tilde{\!\!\n J}} \n p_S \n{\tilde{J}} 
P_L \n k \right\} \no\\   
& & \times\Big\{\mbox{Tr} [W_{j \ell} \n k'\hat{W}_{\ell n} P_L\n k]
+\mbox{Tr}[M_{j \ell} \n k' \hat{M}_{\ell n} P_L\n k]-
m_\ell \mbox{Tr}[(W_{j \ell} \hat{M}_{\ell n} + 
M_{j \ell} \hat{W}_{\ell n}) 
P_L\n k]\Big\}\, .\no\\
\ea
After omitting the production process, the first two terms 
of Eq. (\ref{eq3}) represent the
detection of the antineutrino via the pure weak and the pure electromagnetic
interaction, respectively. If neutrino mixing is neglected these
terms give rise to the well-known cross section (\ref{wq}). The third term 
in (\ref{eq3})
gives the interference terms between weak and electromagnetic interaction
which are proportional to the masses of the final neutrinos. 

It remains to consider the last two terms in Eq. (\ref{sa2}) which depend on 
the masses
of the intermediate neutrinos. It turns out that with the method employed
before factorization cannot be achieved in this case.
Nevertheless, factorization is obtained after
integration over the momentum $k'$ of the outgoing neutrino.
Two points seem to be operative for this goal:
\begin{itemize}
\item rotational invariance around the axis neutrino 
source -- neutrino detection point defined by $\vec{k}$,
\item detector electron in its rest frame.
\end{itemize}
To perform an averaging according to the first of these points
we define an
orthonormal basis with the three vectors $\vec{n}$, $\vec{a}$ and
$\vec{b}$ where
\be
\vec{n}\equiv\frac{\vec{k}}{E_\nu} \, .
\ee
Then we can write
\be\label{phi}
\frac{\vec{k'}}{E'_\nu}=\cos\!\gamma\,\,\vec{n}+\sin\!\gamma
\,\,\vec{m}(\phi)
\ee
with
\be
\vec{m}(\phi)\equiv \cos\!\phi\,\,\vec{a}+\sin\!\phi\,\,\vec{b}\; .
\ee
With this notation we obtain in the rest frame of the detector electron
the relation
 \be\label{intk}
\frac{1}{2\pi}\int\limits_0^{2\pi}\n k'd\phi=\frac{E'_\nu}{E_\nu}\cos\!\gamma
\n k + \frac{E'_\nu}{m_e}(1-\cos\!\gamma)\n p
\ee
where
\be
1-\cos\!\gamma=\frac{m_eT}{E_\nu E'_\nu}\, .
\ee
Note that also $\vec{p\,'}$ contains the angle $\phi$ because
$\vec{p\,'} = \vec{k}-\vec{k}'$.
Using Eq. (\ref{intk}), after some algebra also in the 
last two terms of Eq. (\ref{sa2}) neutrino production can be factorized
from subsequent scattering. Thus the production process can be totally
separated off and also the weak--electromagnetic interference terms can
be written as a cross section for scattering alone.

Finally, the elastic differential antineutrino--electron cross section
in the laboratory frame of the electron is given by
\be\label{s}
\frac {d \sigma}{d T} = \frac{d \sigma_w}{d T}+ \frac {d\sigma_m}
{dT}+\frac{d \sigma_{wm}}{d T}
\ee
with
\ba\label{sw}
\frac{d\sigma_w}{dT}&=&\frac{d\sigma_w(\bar\nu_\mu e^-)}{dT}+ 
\Big|\sum_j e^{iq_jL}|U_{ej}|^2\Big|^2
\left(\frac{d\sigma_w(\bar\nu_e e^-)}{dT}-\frac{d\sigma_w
(\bar\nu_\mu e^-)}{dT}\right)\no\\
&=&\frac{m_eG_F^2}{2\pi}\bigg\{(g_V-g_A)^2+(g_V+g_A)^2(1-\frac{T}{E_\nu})^2
-m_e\frac{T}{E_\nu^2}(g_V^2-g_A^2)\no\\
&&+\bigg|\sum_j e^{iq_jL}|U_{ej}|^2\bigg|^2 
\Big[ 4(1-\frac{T}{E_\nu})^2(1+g_V+g_A) 
-2m_e\frac{T}{E_\nu^2}(g_V-g_A) \Big] \bigg\}\;,
\ea
\be\label{sm}
\frac{d\sigma_m}{dT}=\frac{\alpha^2\pi}{m_e^2}\sum_\ell\bigg|\sum_j e^{iq_jL}
U_{ej}(\mu'_{j\ell}+id'_{j\ell})\bigg|^2\;
(\frac{1}{T}-\frac{1}{E_\nu})
\ee
and
\ba\label{swm}
\frac{d\sigma_{wm}}{dT}&=&\frac{\alpha G_F}{\sqrt{2}E_\nu m_e}\;
\mbox{Re}\bigg\{
\sum_{j,n,\ell}e^{i(q_j-q_n)L}U_{ej}U^*_{en}\no\\
& & \times \bigg( m_\ell(\mu'_{j\ell}+id'_{j\ell})
\Big[ (\frac{m_e}{E_\nu}-\frac{T}{E_\nu})
Z^{V*}_{n\ell}+(2-\frac{T}{E_\nu})Z^{A*}_{n\ell} \Big] \no\\
& & + m_j(\mu'_{j\ell}-id'_{j\ell})
\Big[ (\frac{m_e T}{2E_\nu^2}-1)Z^{V*}_{n\ell}
+(\frac{m_e T}{2E_\nu^2}+1)Z^{A*}_{n\ell} \Big] \bigg)\bigg\}\;. 
\ea
In Eq. (\ref{sw}), $d \sigma_w(\bar\nu_e e^-)/dT$ is identical with the
expression given in Eq. (\ref{wq1}), the elastic $\bar\nu_e e^-$ cross
section without effects of neutrino oscillations.

If neutrino mixing is neglected, i.e. $U$ is the unit matrix,
Eqs. (\ref{sw}) and (\ref{sm}) reproduce the cross sections (\ref{wq1})
and (\ref{wq2}).
In contrast to the pure weak and electromagnetic contributions in
Eq. (\ref{s}), the interference terms (\ref{swm}) depend explicitely
on the neutrino masses. This linear dependence on $m_\nu$ is obviously
needed to provide the necessary helicity flip to make interference between
the weak amplitude (no helicity flip) and the amplitude for magnetic and
electric dipole moments (helicity flip) possible. 
The terms in the second line of
Eq. (\ref{swm}) are proportional to the masses of 
the final state neutrinos whereas
the terms in the third line contain the masses of the incident
neutrinos. In the latter terms the details of the neutrino production
mechanism enter. Thus in the interference
cross section the information that the neutrinos are produced 
by a static $\beta$-decay source is
incorporated. The validity of Eq. (\ref{swm}) for other 
production mechanisms is not guaranteed.

\section{Majorana Neutrinos}

If neutrinos are Majorana particles 
it is usual to define the magnetic (electric dipole) moment part
of the electromagnetic interaction 
Lagrangian (\ref{Lagm}) with an additional factor 1/2.
The neutrino mass eigenfields now fulfill the
Majorana condition $(\nu_i)^c=\nu_i$. Moreover, the matrices $(\mu_{jk})$ and
$(d_{jk})$ are antisymmetric which means that the neutrinos possess only 
transition moments.

Since in the Majorana case also $\nu_j \nu_j^T$ can be Wick-contracted
the above mentioned factor 1/2 is cancelled and an
additional weak term $\mathcal{A}_\ell^M$ occurs in the amplitude (\ref{amp}):
\ba
\label{ampM}
\mathcal{A}_\ell^M & = &
\sum_j \tilde J_\lambda\bar u_{eS}\gamma^\lambda P_L(-\n k_j
+m_j)e^{iq_jL} U_{ej}\no\\
& & \times (-1) \sqrt{2} G_F \gamma^\mu P_R v(k'_\ell)\:
\bar u_e(p')\gamma_\mu
(Z^V_{\ell j}-Z^A_{\ell j}\gamma_5)\tilde\psi_e\; .
\ea
This additional amplitude is proportional to $m_j$ and generates
in our approximation only one additional
term in the $\bar\nu_e e^-$ cross section which is a
weak--electromagnetic interference term. It is given by
\ba
\frac{d\sigma^M_{wm}}{dT} & = & \frac{\alpha G_F}
 {\sqrt 2E_\nu m_e}\mbox{Re} 
\bigg\{ \sum_{j,n,\ell}e^{i(q_j-q_n)L}U_{en}^*U_{ej}m_n
 (\mu'_{j\ell}+id'_{j\ell})\no\\
& & \times \Big[ (\frac{m_eT}{2E^2_\nu}+\frac{T}
 {E_\nu}-\frac{m_e}{E_\nu}-1)Z^{V*}_{\ell n} 
+(1-\frac{T}{E_\nu}
-\frac{m_eT}{2E^2_\nu})Z^{A*}_{\ell n} \Big] \bigg\}  \label{swmm}
\ea
and has to be added to Eq. (\ref{swm}) in the Majorana case.

\section{Discussion}

In this paper we only consider limits on magnetic moments of
neutrinos derived in terrestrial experiments. It should be kept in
mind, however, that cosmological and astrophysical considerations
result usually in more stringent limits 
(for reviews see e.g. Refs. \cite{munu2,cos}) although with assumptions
beyond those made in terrestrial scattering experiments.

The pure weak cross section Eq. (\ref{sw}) with neutrino  oscillations
incorporated is also relevant in the KAMIOKANDE experiment for the
detection of solar \cite{kams} and atmospheric \cite{kama} neutrino fluxes.

The positive result of the LSND experiment \cite{LSND} for $\bar
\nu_\mu \to \bar \nu_e$ transitions together with the negative results
of all other experiments on this channel, in particular, the
experiments of Ref. \cite{app}, and the negative
result of the Bugey $\bar \nu_e$ disappearance experiment \cite{bugey}
restricts the neutrino mass-squared difference $\Delta m^2$
responsible for this transition to $0.3 \lesssim \Delta m^2 \lesssim
2.2 \; \mathrm{eV}^2$ (see also Ref. \cite{BGG}). Since the neutrino
oscillation length is given by $\ell_{\mathrm{osc}} \approx 2.5 \,
\mathrm{m} \, \times (E_\nu/ 1\, \mathrm{MeV})(1\,
\mathrm{eV}^2/\Delta m^2)$ all reactor experiments are at a distance
$L \gtrsim l_{\mathrm{osc}}$ from the reactor core 
where terms in the $\bar \nu_e$ flux are
more or less averaged out. Thus the $\bar \nu_e$ flux is reduced by a
factor $1-\frac{1}{2} \sin^2 2\theta$ with $10^{-3} \lesssim \sin^2
2\theta \lesssim 4 \times 10^{-2}$ determined by the LSND experiment
and the above range of $\Delta m^2$. The transition probability is
given by  $\sin^2 2\theta \sin^2 (\pi L/l_{\mathrm{osc}})$ 
in a two-flavour oscillation formalism. 
This flux reduction seems to be
too small to be detected by presently planned $\bar \nu_e e^-$ scattering
experiments \cite{munu1,munu2}.

Admitting also sterile neutrinos then Eq. (\ref{sw}) gets modified to
\begin{equation}
\frac{d\sigma_w}{dT} = (1-P_{\bar\nu_e \to \bar\nu_s}(L))
\frac{d\sigma_w(\bar\nu_\mu e^-)}{dT}+ 
P_{\bar\nu_e \to \bar\nu_e}(L)
\left(\frac{d\sigma_w(\bar\nu_e e^-)}{dT}-\frac{d\sigma_w
(\bar\nu_\mu e^-)}{dT}\right)
\end{equation}
with survival $(\alpha = \beta)$ or transition $(\alpha \neq \beta)$
probabilites $P_{\bar\nu_\alpha \to \bar\nu_\beta}$. Here $\alpha$ and
$\beta$ denote not only the flavours $e$, $\mu$ and $\tau$ but also
sterile degrees of freedom $(s)$. Apart from the MUNU experiment, 
all presently operating experiments
with reactor neutrinos have $\bar \nu_e + p \to e^+ + n$ as detection
reaction (see, e.g. Ref. \cite{zuber}) and thus measure the survival
probability $P_{\bar\nu_e \to \bar\nu_e}$. Combining the results of
these experiments with future results from elastic $\bar\nu_e e^-$
scattering and neglecting possible neutrino magnetic (electric dipole)
moments would therefore allow to obtain information on 
$P_{\bar\nu_e \to \bar\nu_s}$, the transition probability for
$\bar\nu_e$ into sterile neutrinos. This situation is similar to the
SNO experiment \cite{SNO} for the solar $\nu_e$ flux. If the dynamical
zero in 
$d\sigma_w(\bar\nu_e e^-)/dT$ \cite{bern} can be exploited
then even elastic $\bar\nu_e e^-$ scattering alone would be sufficient.

In order to discuss the cross section $\sigma_m$ (\ref{sm}) it is
useful to define the oscillation amplitude
\begin{equation}\label{A}
\bar A_{\alpha \beta}(L) \equiv \sum_j U^*_{\beta j}U_{\alpha j}e^{iq_jL}
\end{equation}
such that the transition or survival probabilities in neutrino oscillations
are given by
\begin{equation}
P_{\bar\nu_\alpha\to\bar\nu_\beta}(L) = 
\left| \bar A_{\alpha \beta}(L) \right|^2 \, .
\end{equation}
With Eq. (\ref{A}) we can rewrite $\sigma_m$ by using
\begin{equation}\label{sum}
\sum_\ell \left| U_{ej} e^{iq_jL}(\mu'_{j\ell}+id'_{j\ell}) \right|^2 =
\sum_\beta \left| \bar A_{e \alpha} \tilde{\mu}_{\alpha \beta}
\right|^2
\quad \mbox{with} \quad 
\tilde{\mu}_{\alpha \beta} \equiv 
U_{\alpha j}(\mu'_{j\ell}+id'_{j\ell})U^*_{\beta \ell} \, .
\end{equation}
It is obvious that $\sigma_m$ is sensitive to neutrino oscillations as
long as $\tilde{\mu}_{\alpha \beta}$, the neutrino ``moment'' 
matrix in flavour space, or, equivalently,
$\mu'_{j\ell}+id'_{j\ell}$ is not proportional to the unit matrix.
In this case, due to conservation of probability, we obtain
\begin{equation}
\sum_\beta \left| \bar A_{e \alpha} \tilde{\mu}_{\alpha \beta}
\right|^2 \propto \sum_\beta \left| \bar A_{e \beta} \right|^2 =
\sum_\beta P_{\bar\nu_e\to\bar\nu_\beta}(L) = 1
\end{equation}
and the dependence of
$\sigma_m$ on $L$ disappears.

However, even if the moment matrix $\tilde{\mu}_{\alpha \beta}$ is not
proportional to the unit matrix it is probably quite difficult to
exploit this property for experiments on neutrino oscillations because 
by order of magnitude we would expect 
$|\tilde{\mu}_{\alpha \beta}| \sim |\tilde{\mu}_{\beta \alpha}| \lesssim 
10^{-10}-10^{-9}$ \cite{derbin,krakauer} for 
$\alpha = e, \mu$ and $\beta = e, \mu, \tau$. 
For the $\tau$ neutrino the limit is 
$\mu_{\nu_\tau} < 5.4 \times 10^{-7} \mu_B$ \cite{cooper} 
which translates in our framework 
into $|\tilde{\mu}_{\tau \beta}| \sim |\tilde{\mu}_{\beta \tau}| \lesssim 
10^{-6}$. Consequently, only $\tilde{\mu}_{\tau \tau}$ could be as large 
as the limit from Ref. \cite{cooper}. In addition, there are no terrestrial
limits on the magnetic moments and electric dipole moments for neutrino 
degrees of freedom which are sterile. If there are neutrino oscillations 
into the $\tau$ neutrino and/or sterile neutrinos these transitions 
could be enhanced by corresponding large magnetic moments in the 
electromagnetic cross section Eq. (\ref{sm}). One should, however, 
take into account that then the same electromagnetic cross section is also
present in the KAMIOKANDE experiment measuring the solar 
\cite{mourao} and atmospheric neutrino fluxes 
in which neutrinos with large magnetic moments
of order $10^{-6}$ can therefore not be present in large quantities.
It has been noted \cite{bern} that the dynamical
zero in the $\bar\nu_e e^-$ scattering can be used to discover new
physics and, in particular, to study neutrino magnetic moments.

The order of magnitude of the 
interference term $\sigma_{wm}$, Eqs. (\ref{swm}) and (\ref{swmm}),
is estimated by considering the factors appearing in it which leads
to
\begin{equation}\label{oswm}
\frac{\alpha G_F m_\nu \mu'_\nu}{\sqrt{2} m_e} (\hbar c)^2 \approx
4.59 \times 10^{-51} \:\mathrm{cm}^2\: \times 
(m_\nu / 1 \mathrm{eV})\,(\mu'_\nu/10^{-10})
\end{equation}
where $\mu'_\nu$ denotes a generic neutrino magnetic moment or
electric dipole moment in units of $\mu_B$. Note that in
Eq. (\ref{oswm}) we have put $m_e$ in the denominator rather than
$E_\nu$ as expected from a superficial look at Eq. (\ref{swm}) or
(\ref{swmm}). The reason is that a factor $1/E_\nu$ in
Eq. (\ref{oswm}) would suggest that $\sigma_{wm}$ rises for $E_\nu \to
0$. This is not the case because the upper boundary of $T$ goes to
zero like $E_\nu^2$ in this limit:
\begin{equation}
0 \leq T \leq T_{\mathrm{max}} = \frac{2E_\nu^2}{2E_\nu + m_e} \, .
\end{equation}
Eq. (\ref{oswm}) correctly indicates the behaviour of the integrated
interference cross section $\sigma_{wm}$ for $E_\nu \to 0$ and 
$E_\nu \to \infty$.

Eq. (\ref{oswm}) has to be compared with analogous quantities
\begin{equation}
\frac{G_F^2 m_e E_\nu}{2\pi} (\hbar c)^2 \approx 
4.31 \times 10^{-45} \:\mathrm{cm}^2 \times (E_\nu/1\, \mathrm{MeV})
\end{equation}
and
\begin{equation}
\frac{\alpha^2 \pi \mu'^2_\nu}{m_e^2} (\hbar c)^2 \approx 2.49 \times
10^{-45} \: \mathrm{cm}^2 \times (\mu'_\nu/10^{-10})^2
\end{equation}
determining the orders of magnitude of $\sigma_w$ and $\sigma_m$,
respectively. Although in $\sigma_{wm}$ the neutrino moments appear
only linearly, this term is about six orders of magnitude smaller than
$\sigma_w$ and $\sigma_m$ for $m_\nu \sim 1$ eV. 
The latter two cross sections are of the
same order of magnitude for $E_\nu \sim 1$ MeV and $\mu'_\nu \sim 10^{-10}$
which is the reason that in elastic 
neutrino--electron scattering it is possible that stringent bounds on
neutrino moments can be obtained experimentally. The interference term
$\sigma_{wm}$ can only be of similar order of magnitude for neutrino
oscillations into degrees of freedom with very large magnetic moments
as discussed before \emph{and} neutrino masses e.g. in the keV range
which are not favoured by present hints for neutrino oscillations.

In conclusion, we have calculated the weak--electromagnetic
interference terms for elastic $\bar \nu_e e^-$ scattering of neutrinos
with masses, mixing and magnetic and/or electric dipole
moments. In lowest order these terms are linear in the neutrino
masses. To study such interference terms it is necessary to know the
initial neutrino state in $\bar\nu_e e^-$ scattering, which is a
superposition of neutrino mass eigenstates and corresponding helicity
states, to the same precision,
i.e. up to terms linear in the neutrino masses. To this end we have
investigated neutrino production and scattering as a single combined
process. Though naturally in this way production and scattering
processes are entangled it is easy to see that for pure weak and
electromagnetic antineutrino--electron scattering, the production and
scattering processes factorize as expected at order $m_\nu^0$ . 
This is not the case anymore at the next non-trivial 
order $m_\nu^2$. (There are no contributions linear in $m_\nu$.) As for
weak--electromagnetic interference which occurs at order $m_\nu$ we could
show that for the usual
experimental set-up of experiments with reactor neutrinos (neutrino
source and electron target at rest, rotational invariance around the
axis source -- detector) production factorizes from scattering as
well. In this way we obtained
a weak--electromagnetic interference cross section which adds to the
pure weak and electromagnetic $\bar\nu_e e^-$ scattering cross
sections. However, it
turns out that for neutrino masses of order 1 eV and magnetic moments
of order $10^{-10} \mu_B$ this interference cross section is approximately six
orders of magnitude suppressed compared to the pure weak and
electromagnetic cross sections for neutrino energies $\sim$ 1
MeV. Thus under the usual assumptions one can neglect this term in
elastic neutrino--electron scattering. We have also commented on the
possibility of using the electromagnetic neutrino cross section as a
means for the investigation of neutrino oscillations. This could be
promising for a large magnetic moment (or electric dipole moment) of the
$\tau$ neutrino.

\acknowledgments
We thank S. M. Bilenky for drawing our attention to the problem
studied in this paper.

\end{document}